\documentstyle{article}

\makeatletter

\expandafter\ifx\csname documentclass\endcsname\relax
  \def\@jscquest{{\bf ?}}
  \long\def\@firstofone#1{#1}
  \let\G@refundefinedtrue\relax
  \let\@latex@warning\@warning
\else
  \def\@jscquest{\leavevmode\mbox{\reset@font\bfseries ?}}
\fi
 
\let\@internalcite\cite
\def\@normalcitesep{;\penalty-\@m\ }%
\def\@ccitesep{, }%
\def\cite{\let\@citesep\@normalcitesep
 \def\@cite##1##2{(\nobreak\hskip 0in{##1\if@tempswa , ##2\fi})}%
 \def\citeauthoryear##1##2{##1,\penalty\@m\ ##2}\@internalcite}
\def\ccite{\let\@citesep\@ccitesep
 \def\@cite##1##2{{##1\if@tempswa , ##2\fi}}%
 \def\citeauthoryear##1##2{##1\penalty\@m\ (##2)}\@internalcite}
 
\def\citeauthor#1{%
 \def\citeauthoryear##1##2{##1}\@citedata{#1}}
\def\citeyear#1{%
 \def\citeauthoryear##1##2{##2}\@citedata{#1}}
 
\def\@citedata#1{%
 \if@filesw\immediate\write\@auxout{\string\citation{#1}}\fi
 \@ifundefined{b@#1}{\@jscquest	
   \G@refundefinedtrue\@latex@warning
   {Citation `#1' on page \thepage \space undefined}}%
  {\csname b@#1\endcsname}}
 
\def\@citex[#1]#2{%
  \let\@citea\@empty
  \@cite{\@for\@citeb:=#2\do
    {\@citea\let\@citea\@citesep
     \edef\@citeb{\expandafter\@firstofone\@citeb}%
     \if@filesw\immediate\write\@auxout{\string\citation{\@citeb}}\fi
     \@ifundefined{b@\@citeb}{\@jscquest	
       \G@refundefinedtrue\@latex@warning
       {Citation `\@citeb' on page \thepage \space undefined}}%
     {\csname b@\@citeb\endcsname}}}{#1}}

\def\@biblabel#1{}
\def\thebibliography#1{\section*{References}
  \addcontentsline{toc}{section}{References}
  \list{}{\labelwidth\z@
    \leftmargin 1.5pc
    \itemindent-\leftmargin}
    \footnotesize
    \parindent\z@
    \parskip\z@ plus .1pt\relax
    \def\newblock{\hskip .11em plus .33em minus .07em}
    \sloppy\clubpenalty4000\widowpenalty4000
    \sfcode`\.=1000\relax}

\@addtoreset{equation}{section}

\makeatother

\newcommand\bm[1]{\mbox{\boldmath $#1$}}
\font\SYM=msbm10 
\def\Real{\hbox{\SYM R}}
\newcommand{\smfrac}[2]{{\textstyle{#1\over#2}}}
\def\half{\smfrac{1}{2}}
\def\cosech{\mathop{\rm cosech}\nolimits}
\def\sech{\mathop{\rm sech}\nolimits}

\renewcommand{\d}{{\rm d}}
\newcommand{\e}{{\rm e}}
\makeatletter

\@addtoreset{equation}{section}

\makeatother

\begin{document}

\title{Hypersurface-orthogonal generators of an orthogonally transitive
 transitive $G_2I$, topological identifications, and axially
 and cylindrically symmetric spacetimes}

\author{M A H MacCallum \\
School of Mathematical Sciences, \\
Queen Mary and Westfield College,\\
London E1 4NS, UK.\\
E-mail address: M.A.H.MacCallum@qmw.ac.uk}

\maketitle

\begin{abstract}
A criterion given by \ccite{CasMac90} for the existence of (locally)
hypersurface-orthogonal generators of an orthogon\-ally-transitive
two-parameter Abelian group of motions (a $G_2I$) in spacetime is
re-expressed as a test for linear dependence with constant
coefficients between the three components of the metric in the orbits
in canonical coordinates. In general, it is shown that such a relation
implies that the metric is locally diagonalizable in canonical
coordinates, or has a null Killing vector, or can locally be written
in a generalized form of the `windmill' solutions characterized by
McIntosh. If the orbits of the $G_2I$ have cylindrical or toroidal
topology and a periodic coordinate is used, these metric forms cannot
in general be realized globally as they would conflict with the
topological identification. The geometry then has additional essential
parameters, which specify the topological identification. The physical
significance of these parameters is shown by their appearance in
global holonomy and by examples of exterior solutions where they have
been related to characteristics of physical sources. These results
lead to some remarks about the definition of cylindrical symmetry.

Keywords: cylindrical symmetry, stationary, Killing vectors, topology,
holonomy
\end{abstract}


\section{Introduction}

This paper considers spacetime metrics with a pair of
orthogonally-transitive commuting Killing vectors (KVs), generating a
group of type $G_2I$ (Abelian). The most often considered cases are
those in which it is possible to write the metric locally in the form
(see \ccite{KraSteMac80}, pages 194 and 220)
\begin{equation}
\label{SS}
\d s^2 = \e^{-2U}[\e^{2k}(\d \rho^2 - \d t^2) +\rho^2 \d \phi^2] +
\e^{2U}(\d z+A\,\d \phi)^2
\end{equation}
where $U$, $k$ and $A$ are real functions of $t$
and $\rho$ only, if the $G_2$ acts on spacelike surfaces, or
\begin{equation}
\label{TS}
\d s^2 = \e^{-2U}[\e^{2k}(\d \rho^2 + \d z^2) +\rho^2 \d \phi^2] -
\e^{2U}(\d t+A\,\d \phi)^2,
\end{equation}
where $U$, $k$ and $A$ are real functions of $z$ and $\rho$ only, if
the $G_2$ acts on timelike surfaces. (These forms, or the special
cases considered by \ccite{Hof69} which will not be discussed here,
are valid for vacuum and some other energy-momentum tensors. In
certain circumstances, such as the existence of regular axis points,
the orthogonal transitivity of the $G_2$ can be proved rather than
assumed, but I do not pursue that point here.) The two forms just
given can both be written as
\begin{equation}
\label{genmet}
\d s^2 = \e^{-2U}[\e^{2k}(\d \rho^2 - \zeta \d Y^2) +\rho^2 \d \phi^2]
+ \zeta \e^{2U}(\d X+A\,\d \phi)^2
\end{equation}
where $\zeta = \pm 1$, $U$, $k$ and $A$ are functions of $Y$ and
$\rho$ only, and, if $\zeta = 1$, then $Y=t$ and $X=z$, while if $\zeta
= -1$, then $Y=z$, $X=t$. The KVs $\bm{\xi} = \partial/\partial X$  and
$\bm{\eta}=\partial/\partial \phi$ form a basis of the generators of the
$G_2$.

One may also write the metric, in slightly more general coordinates
which no longer restrict the energy-momentum, as
\begin{equation}
\label{Lewis}
\d s^2 = \e^{\mu}\d r^2 - \zeta \e^{\nu}\d Y^2 + \ell \, \d \phi^2 +
2m \, \d \phi \, \d X + \zeta f\, \d X^2
\end{equation}
where the metric components depend only on $r$ and $Y$. The form
(\ref{genmet}) is the case where $r=\rho$ and $\mu=\nu=U$; the metric
coefficients of the cases above are included in this form, with
corresponding values $\ell \equiv \chi
= (\rho^2 + \zeta (A\e^{2U})^2)\e^{-2U} $ ($\chi$ being the notation
of \ccite{CasMac90}), $m = \zeta A\e^{2U}$, and $f=\e^{2U}$. One can
of course choose new mutually orthogonal coordinates $\hat{r} =
\hat{r}(r,\,Y)$, $\hat{Y} = \hat{Y}(r,\,Y)$ without altering the
general form (\ref{Lewis}). I shall refer to equation (\ref{Lewis})
for brevity as the Lewis form, cf.\ \ccite{SilHerPai95a}, since it was
used by \ccite{Lew32} with $\zeta = -1$ (equation (1.1) there). I
define $\rho$ in all cases by $\rho^2 = f\ell-\zeta m^2$.

When, for example, investigating whether a gravitational wave solution
is linearly polarized, it is of interest to know whether among the
generators of the $G_2$ there is a hypersurface-orthogonal Killing
vector\footnote{If the $G_2$ under consideration is not the maximal
group of motions, there may be an HSO KV which is not one of its
generators. This happens, for example, in some stationary
cylindrically symmetric metrics discussed later in this paper, but the
possibility will not be discussed fully here.}. In
\ccite{CasMac90}, it was shown that if neither $\bm{\xi}$ nor
$\bm{\eta}$ are themselves hypersurface-orthogonal (HSO), then, among
the generators of the $G_2$, there is one which is HSO and non-null if
and only if an equation
\begin{equation}
\label{crit1}
{1 \over C} = -\zeta {{B \chi + \zeta A \e^{2U}} \over
{\e^{2U}(1+B A)}}
\end{equation}
holds, with distinct real constants $B$ and $C$. In fact, if there is
one, there must be a second, orthogonal
to the first, and the two HSO KVs can be taken to be $\bm{\xi}+B
\bm{\eta}$ and $\bm{\xi} + C\bm{\eta}$. Equation (\ref{crit1}) can be stated in
a simpler form as the requirement that the metric coefficients of the
Lewis form obey a linear relation
\begin{equation}
\label{crit2}
\zeta f(r,\,Y)+ a m(r,\,Y) + b \ell(r,\,Y) = 0.
\end{equation}
with constant coefficients $a = (B +C)$ and $b= B C$. Despite the
simplicity of this form of the criterion, it did not appear in a
general form in any earlier paper known to me. (However, similar
statements in the context of cylindrically symmetric stationary vacuum
and electrovacuum solutions were given by \ccite{ArbSom73} and
\ccite{SomSan78}, and, for the case of spacelike orbits, a form of the
general criterion has just appeared as equation (9) in
\ccite{MarSen97}: I thank J.M.M. Senovilla for drawing the latter to
my attention.)

In this form it is often immediately apparent whether the criterion is
satisfied or not, by inspecting whether or not the functions $f$, $m$
and $\ell$ contain only two functions of $r$ and $Y$ linearly
independent over $\Real$. When the criterion is hard to test, due to
the complexity of the functions being considered, one can find the
coefficients in equation (\ref{crit2}) by evaluation at specific $r$
and $Y$: consistency of the equations at three (or more) such
evaluation points is required, and of course the resulting $B$ and $C$
must be valid at all points. Clearly, if the criterion (\ref{crit2})
is satisfied, $B$ and $C$ will simply be the roots of
\begin{equation}
\label{quadr}
w^2- a w + b = 0.
\end{equation}

This prompts the question: what if the functions $f$, $m$ and $\ell$
are linearly dependent but the associated quadratic does not have
distinct real roots? In Section 2 I show that if the functions $f$,
$m$ and $\ell$ are linearly dependent with constant coefficients
(necessarily real), then either there is a pair of non-null HSO KVs
(and the metric is locally diagonalizable\footnote{Throughout this
paper, `diagonalizable' means `diagonalizable in a holonomic frame two
of whose coordinates are ignorable', i.e.\ in the form (\ref{Lewis})
with $m=0$. Two-dimensional metrics, being conformally flat, can
always be diagonalized, but not necessarily in a way consistent with
the form (\ref{Lewis}). Also, metrics of any dimension can be
diagonalized in non-holonomic tetrads, e.g.\ orthonormal tetrads. In
older papers where this fact was used, confusion was sometimes
created, at least for a modern reader, by the practice of notating the
basis one-forms as if they were coordinate differentials, for example
in equations (3.1) of \ccite{Lew32}.}), or there is a null HSO KV,
which one can regard as the limiting case where the two non-null HSO
KVs coincide, or the metric can be expressed in a generalization of
the McIntosh `windmill' form for vacuum solutions \cite{McI92}. Note
that this formulation avoids the difficulties of expressing, within
some version of the above formulae, the cases where one or both
Killing vectors have already been aligned with hypersurface-orthogonal
ones. The last of the three cases can be viewed as having a pair of
complex conjugate hypersurface-orthogonal KVs. For certain vacuum
metrics, this result is implicit in section 18.4 of
\ccite{KraSteMac80} and references cited therein (see also
\ccite{SomSan78}) but again the general statement appears to be new.

These considerations are purely local. They will therefore be
consistent with the results one can obtain from the local
characterization of the metric in terms of Cartan scalars
\cite{Kar80,MacSke94,PaiRebMac93,SilHerPai95a}. However, the metric
may have parameters which are important globally but do not appear in
the Cartan scalars. This is considered further in section 3 for the
case of axisymmetric metrics, where the coordinate suggestively named
$\phi$ is normally assumed to be periodic. Choosing $\phi$ to have the
usual period $2\pi$, one finds that the locally equivalent geometries
with distinct identifications in the $G_2$ orbits, and the choices of
ignorable coordinates consistent with them, can be specified by three
parameters. On careful consideration of possible coordinate choices it
turns out that only two of the three parameters are essential, in that
they cannot be removed by permissible changes of coordinates. (In this
paper, `essential' refers only to unique characterization of the geometry,
not to any other purpose.)

The three parameters can be interpreted, in terms of the
identification, as follows.  If the metric in the $(\phi,\,X)$ plane
takes a standard form (as described in Section 2) in coordinates
$(\hat{\phi},\, \hat{X})$, two of the three parameters specify a point
$(\hat{\phi}_0,\,\hat{X}_0)$ which is to be identified with the
origin, and the third (the inessential one) specifies the direction of
parallel lines through the origin and $(\hat{\phi}_0,\,\hat{X}_0)$
along which the identification is to take place. To understand the
relation between locally equivalent but globally inequivalent
axisymmetric metrics with the Lewis form, one can imagine unrolling
the cylinder $0 \leq \phi \leq 2\pi$, $-\infty < X < \infty$ into a
plane and then rolling it up again in a different way.

From a physical point of view one would wish to associate the
additional essential parameters with curvature, and this is discussed
in section 4. The parameters cannot change the values of the Cartan
scalars defined by the Riemann tensor and its derivatives at a point,
and this directs attention to the possible global holonomy found by
taking suitable closed curves, i.e.\ the `gravitational Aharonov-Bohm
effect' of \ccite{Mar59} and \ccite{Sta82}. Stachel points out that
for a given metric form, the linear holonomy will depend only on that
metric's curvature unless the region in which it is defined is not
simply connected (has non-zero first Betti number).

Stachel does not give explicit general formulae for this holonomy
(though he calculates some specific results in \ccite{Sta84}). These are
derived in Section 4. In general the two essential parameters of the
identification do appear in these formulae. The metrics studied may be
matched to some regular source in an interior region with a different
metric; in this case, the holonomy not due to the curvature of the
exterior is due to the curvature in the region occupied by the
source. If the solutions are valid for all points except some singular
axis (a concept which as yet lacks precise definition, see
\ccite{MarSen95}), the holonomy may be ascribed to singular sources on
the axis. \ccite{Vic87} and \ccite{Tod94} have considered the linear
and affine holonomy of spacetimes constructed by identifications on
flat space, and Tod has remarked that the non-zero affine holonomy
could be considered to arise from a distributional torsion on the axis
and can also be viewed as an example where a non-local effect of a
symmetric connection mimics a local effect of a non-symmetric
connection. \ccite{WilCla96} considered the general behaviour of
holonomy at an axis.

When there is a definite interior matched to the region being
considered, all the essential parameters will be fixed by the
matching, using the coordinate-free Darmois form of the boundary
conditions. However, if one wishes (a) to write the matching
conditions in the equivalent Lichnerowicz form, which differs in that
`admissible coordinates', smooth across the boundary and always
existent if the Darmois conditions are true and the spacetime is
smooth enough (see \ccite{BonVic81}), must be used, and (b) to tie the
coordinates in the inner region to physical characteristics there,
then some or all of the inessential parameters in the exterior may
also be fixed. For vacua exterior to rotating shells or fluids, such matchings
have been considered by, e.g.,
\ccite{Bon80,Sta84,SilHerPai95,SilHerPai95a,BonMacSan97}.

These calculations prompt a criticism of the definition of cylindrical symmetry
given in \ccite{KraSteMac80}, which is too restrictive.

\section{The general form of the criterion}

\noindent{\bf Proposition.} {\em 
If the metric components $f$, $m$ and $\ell$ in the Lewis metric form {\rm
(\ref{Lewis})} are linearly dependent over $\Real$, then by a
homogeneous linear transformation of the coordinates
$(\phi,\,X)$ with constant coefficients,\\
(i) the metric can be put in a locally diagonal form,
\begin{equation}
\label{diag}
\d s^2 = \e^{2V}(\d r^2 - \zeta \d Y^2) + \ell \, \d \hat{\phi}^2 +
\zeta f \,\d \hat{X}^2,
\end{equation}
where $V$, $\ell$ and $f$ are functions of $r$ and $Y$ only, and it
is manifest that there are two hypersurface-orthogonal Killing vectors, or\\
(ii) the metric has a null Killing vector and can locally be put in the form
\begin{equation}
\label{nullKV}
\d s^2 = \e^{2V}(\d r^2 + \d z^2) + \ell \,\d \hat{\phi}^2 + 2m\,
\d \hat{\phi}\, \d \hat{t},
\end{equation}
where $V$, $\ell$ and $m$ are functions of $r$ and $z$ only, or\\
(iii) the metric can locally be put in the form
\begin{equation}
\label{windmill}
\d s^2 = \e^{2V}(\d r^2 + \d z^2) + \rho [(\cos \psi\, \d \hat{\phi}
+ \sin \psi \, \d \hat{X})^2 - (- \sin \psi\, \d \hat{\phi} + \cos
\psi \, \d \hat{X})^2]
\end{equation}
where $V$, $\rho $ and $\psi$ are functions of $r$ and $z$ related to a
Lewis form {\rm (\ref{Lewis})} with $\zeta = -1$ by $2V=\mu=\nu$,
$\rho ^2=f\ell  + m^2$, $\ell = f = \rho  \cos(2\psi)$, $m=\rho  \sin
(2\psi)$. This last case has a pair of complex conjugate
hypersurface-orthogonal Killing vectors.
} 
\vspace{\baselineskip}

{\bf Remark.} The first of these possibilities includes the well-known
static vacua associated with the names of Weyl, Kasner and
Levi-Civita, and Lewis's first solution: note that if $\zeta = -1$ one
of the KVs $\partial/\partial \phi$ and $\partial/\partial X$ in
equation (\ref{diag}) must be timelike. The second possibility
includes the limiting case of Lewis's solutions given explicitly by
van Stockum though not in the form (\ref{nullKV}) (see equations
(11.1) of \ccite{Sto37}, and the examples below), and the third
includes Lewis's second solution (a complex continuation of the first)
and can be seen to be a generalized form of the vacuum `windmill'
solutions discussed by \ccite{McI92}.
\vspace{\baselineskip}

{\bf Proof.}
To prove the proposition, I first consider those cases where the
quadratic (\ref{quadr}) is in some way degenerate. The linear
dependence may arise simply because one of the three functions
vanishes. If $m=0$, clearly the metric is already in the form required for
case (i). If $f =0$, then $\partial/\partial X$ is a null
KV, $\zeta = -1$ and $m \neq 0$ in order to have a
non-degenerate metric of the correct signature, and the
metric can be written in the form (\ref{nullKV}); if in addition $\ell/m$
is constant, $\ell$ can be set to zero by a further constant
linear transformation of $\phi$ and $t$. If $\ell=0$ then
$\partial/\partial \phi$ is a null KV, and the form (\ref{nullKV}) can be
obtained by exchanging $\phi$ and $X=t$.

Next I consider the case where only two of the functions appear with
non-zero coefficients in the linear dependence.  If the linear
dependence takes the form $m=c\ell$ with $c \neq 0$, then $\hat{\phi}
= \phi+c X$ diagonalizes the metric (note that $\zeta f$ cannot be
$c^2 \ell$ or the metric is degenerate), and similarly if $m=cf$ with
$c\neq 0$, $\hat{X} = X+c\phi$ diagonalizes the metric. The case
where $f=c \ell$ is included in the general discussion of equation
(\ref{crit2}), as the special case where $a=0$: however, dealing with
it separately may help to clarify what happens in the more general
case.

If $f=c\ell$, then by scaling $X$ one can always set $c$ to
sgn($c$). If $\zeta=1$, then $f$, $\ell$ and $f\ell-m^2$ must be
positive in order for the metric to have correct signature, so $c > 0$
and one can take $c=1$. The metric in the $(\phi,\,X)$ plane is then
\[ \ell(\d \phi^2 + \d X^2) + 2m\, \d \phi \,\d X \] and taking
$\hat{X}=X+\phi$, $\hat{\phi}=X-\phi$ (agreeing with the roots $\pm 1$
given by the quadratic (\ref{quadr})) gives the diagonal form \[ \half
[(m+\ell) \d \hat{X}^2 + (\ell -m) \d \hat{\phi}^2] \] in which both
metric coefficients must be positive.  If $\zeta = -1$ and $c < 0$,
one can take $c=-1$ and diagonalize the resulting metric in a similar
way, except that now $m^2 > f\ell$ so one of the final metric
coefficients is negative. If $\zeta=-1$ and $c>0$, one can take $c=1$,
and then by defining $\rho $ and $\psi$ as in (iii) above, one can arrive
at the form (\ref{windmill}). In this case the quadratic (\ref{quadr})
has roots $\pm i$, and the metric can be regarded as having complex
conjugate HSO KVs, taking a diagonal form (involving the exponentials
of $\pm 2i\psi$) with respect to the complex coordinates $\phi \pm i
X$.

In the general case, where all three of $f$, $\ell$ and $m$ enter the
linear dependency relation with non-zero coefficients, one can make the
coefficient of $f$ equal to 1, and then inspect the resulting
quadratic (\ref{quadr}). If this has distinct real roots,  case
(i) applies and the coordinate transformation
\begin{equation}
\label{trans}
X = \hat{X} + \hat{\phi}, \qquad \phi = B \hat{X} + C\hat{\phi}
\end{equation}
gives the metric a diagonal form, with $C^2\ell +2 mC + \zeta f$ and
$B^2 \ell +2B m + \zeta f$ as the new values of $\ell$ and $\zeta
f$. The perhaps surprising form of the first part of equation
(\ref{trans}) merely reflects the choice of scaling of $\hat{\phi}$:
the form
\begin{equation}
\label{trans1}
X = \hat{X} + \hat{\phi}/C, \qquad \phi = B \hat{X} + \hat{\phi}.
\end{equation}
might be considered more natural. If the quadratic (\ref{quadr}) has
two equal real roots then $\partial_X + B\partial_\phi$ is a null
Killing vector, and the coordinate transformation $\phi = \hat{\phi} +
B \hat{X}$, $X = \hat{X}$ brings the metric into the form
(\ref{nullKV}). Finally if the quadratic (\ref{quadr}) has two complex
conjugate roots the transformation (\ref{trans}) gives a complex
diagonal form in which the new $\ell$ and $\zeta f$ are complex
conjugates. Writing these as $-\rho  \e^{2i\psi}$ and $-\rho  \e^{-2i\psi}$
respectively, setting $\sqrt{2}\hat{\phi} = u + iv$, $\sqrt{2}\hat{X} =
u - iv$, and following this by renaming $u$ as $\hat{X}$ and $v$ as
$\hat{\phi}$, gives the form (\ref{windmill}). Note that the form of
the metric is much more easily established if one goes via the complex
form than if one makes the corresponding real transformation
directly. However, the overall coordinate transformation is strictly
real: it is just
\begin{equation}
\label{trans1c}
X=\sqrt{2}\hat{X}, \qquad \phi=\sqrt{2}(b_1\hat{X}-b_2 \hat{\phi})
\end{equation}
where $b_1$ and $b_2$ are respectively
the real and imaginary parts of $B=C^\ast$. This completes the proof.

The forms (\ref{diag}) and (\ref{nullKV}) are unaffected by re-scaling
each of the coordinates $\phi$ and $X$ by constant factors, so at most
two of the parameters of the allowed homogeneous linear
transformations of these coordinates have been used. To preserve the
form (\ref{windmill}) the corresponding scalings have to match: this
extra requirement could be removed by suitably putting an additional
constant into the form (\ref{windmill}). However, one would normally
use the scalings to remove as far as possible any inessential
parameters in the metrics (\ref{diag})--(\ref{windmill}), and in later
sections we shall assume that such a scaling has been chosen in order
to standardize the form of the locally equivalent metric considered:
this is consistent with the form (\ref{windmill}).

Thus when the functions $f$, $\ell$ and $m$ are linearly dependent,
the problem of solving the field equations can be simplified by solving the
equations for the metrics (\ref{diag})--(\ref{windmill}) above and then
transforming. I have not attempted to characterize all cases where
this applies. It does follow from the ansatz used in \ccite{Sto37},
section 9, which is that $v_{,D}=\Theta_{,D}(u+v^2)$,
$u_{,D}=\Phi_{,D}(u+v^2)$ where $v=m/\ell$, $u=f/\ell$ and $x^D$ is
either $r$ or $Y$ (the field equations for vacuum give only that
 \[ \sum_D [r u_{,D}/(u+v^2)]_{,D} = 0 = \sum_D [r
v_{,D}/(u+v^2)]_{,D}, \]
which is consistent with but does not imply
the ansatz in general). This ansatz
has the consequence $v=Mu+N$ for constants $M$ and $N$ (or $u$ or $v$
may be zero or infinity, which van Stockum does not mention). For
stationary cylindrically symmetric metrics of the form (\ref{TS})
whose energy-momentum components in the $(\phi,\,t)$ plane are a
multiple of the metric in that plane this is not an ansatz but can be
proved (see, e.g., \ccite{San93}).

It may also be worth noting that in the form (\ref{genmet}), $\rho =
0$ is usually referred to as the axis, though since the geometry may
be singular there, the points may not be a properly-defined part of
the Riemannian manifold (cf.\ \ccite{MarSen95}). If there is such an axis, if
$f$, $m$ and $\ell$ are linearly dependent, and if at the axis $f \neq 0$,
then either $m=0$ (and the metric was diagonal from the start), or the
limiting value of $f/m$ as $\rho \rightarrow 0$ necessarily gives one
of the roots of the (possibly degenerate) quadratic (\ref{quadr}).

To illustrate the result I look again at the examples studied by
\ccite{CasMac90}. Three of these are in fact forms of Lewis's
stationary cylindrically symmetric vacuum solutions.  The
\ccite{ArbSom73} solutions are explicitly derived by a transformation
from a static form (giving the `Weyl class' treated in
\ccite{SilHerPai95a}, so called because they are related by the
a transformation (\ref{trans1}) to solutions in Weyl's class of static
axisymmetric spacetimes), and are thus locally static, i.e.\ fall into
case (i) above. (Whether these should be considered globally static,
as claimed in \ccite{CasMac90}, is discussed below.) These solutions
have
\begin{equation}
\begin{array}{lcl}
f &=& \gamma^2(\exp(2\alpha)-\omega^2 r^2 \exp(-2\alpha)) \\
m &=&  \gamma^2 \omega(\exp(2\alpha) - r^2 \exp(-2\alpha)) \\
\ell &=& \gamma^2 (r^2 \exp(-2\alpha)- \omega^2\exp(2\alpha))
\end{array}
\end{equation}
where $\gamma^2(1-\omega^2)=1$, $\omega$ is constant, and $\alpha$
depends on $r$; from this one can immediately see that $-f+(\omega +
1/\omega)m+\ell=0$, the roots of the quadratic (\ref{quadr}) are
$\omega^{\pm 1}$, and the HSO KVs given in \ccite{CasMac90} (which
invert the construction of \ccite{ArbSom73}) are recovered.

Bonnor's discussion (\citeyear{Bon80}) of the van Stockum solutions
(\citeyear{Sto37}) shows that only in van Stockum's first case are the
exterior solutions locally static. In this case, the functions in
(\ref{Lewis}) in the coordinates used by Bonnor are
\begin{equation}
\label{regionI}
\begin{array}{lcl}
f &=& (r \sinh(\epsilon - \theta) \cosech \epsilon)/R\\
m &=& r \sinh(\epsilon + \theta) \cosech 2\epsilon \\
\ell &=& (r R \sinh(3\epsilon+\theta) \cosech 2\epsilon \sech \epsilon)/2
\end{array}
\end{equation}
where $\theta = \sqrt{1-4a^2R^2} \log (r/R)$, $\tanh \epsilon =
\sqrt{1-4a^2R^2}$ and $a$ and $R$ are constants. It is obvious that
$f$, $m$ and $\ell$ depend linearly on the two functions $r \sinh \theta$
and $r \cosh \theta$. The dependence is
 \[ -Rf - 4 \cosh \epsilon \cosh(2\epsilon) m +4\cosh^2 \epsilon
\,\ell/R = 0 \]
with roots $-(2 \e^{\pm 2\epsilon} \cosh \epsilon)/R$ for the
quadratic (\ref{quadr})
and resulting HSO KVs as in \ccite{Bon80,CasMac90}.

In van Stockum's second case, the one not given explicitly by Lewis,
\begin{equation}
\begin{array}{lcl}
f &=& r/R (1-\log (r/R))\\
m &=& r (1+\log (r/R))/2 \\
\ell &=& r R (3+\log (r/R))/4
\end{array}
\end{equation}
and $-Rf -4m +4\ell/R=0$, giving coincident roots $-2/R$ for the quadratic
(\ref{quadr}) and thus a null KV (as found by \ccite{Bon80,CasMac90});
this is case (ii) above.

Finally, the third case is the same as the first but with all the
hyperbolic functions replaced by the corresponding trigonometric ones,
$\theta = \sqrt{4a^2R^2-1} \log(r/R)$ and \\ $\tan \epsilon =
\sqrt{4a^2R^2-1}$ ($0 \leq \epsilon < \pi/2$): this is equivalent to
the complexified form given by Lewis. The linear dependency is $-Rf - 4
\cos \epsilon \cos(2\epsilon) m +4\cos^2 \epsilon \,\ell/R = 0$, the
roots of the quadratic (\ref{quadr}) are $-(2 \e^{\pm 2i\epsilon} \cos
\epsilon)/R$ and case (iii) holds.

The paper of \ccite{SomTeiWol76} also considered the problem of
stationary cylindrically symmetric vacuum metrics and derived the
linear dependency directly (without first obtaining explicit forms for
$f$, $m$ and $\ell$) in the manner mentioned above. They then concluded
that all such metrics can be diagonalized, but this is only true if
one allows complex coordinates as in case (iii), and overlooks the
case (ii). In the notation of \ccite{SomTeiWol76}, $m=\delta f +
\gamma \ell$ and  cases (ii) and (iii) arise when their $\eta =
\sqrt{1+4\delta\gamma}$ becomes zero or negative.

Lastly, the metrics given by \ccite{PirSafKat86} have $\zeta = 1$,
\begin{eqnarray}
f &=& {{[\alpha^2(1-\lambda_u\lambda_v)^2+(\lambda_u^2+\lambda_v^2)]}
\over {[(\alpha^2\Xi^2 +(\lambda_u-\lambda_v)^2]}}\\
m &=& -{{a\sqrt{\alpha^2-1}[\Xi
(\lambda_v+\lambda_u)^2]}
\over {\sqrt{\lambda_u\lambda_v}[(\alpha^2\Xi^2 +(\lambda_u-\lambda_v)^2]}}\\
\ell&=&(\rho^2 +m^2)/f
\end{eqnarray}
where $a$ and $\alpha = M/a$ are constants,
$\lambda_u=[\sqrt{a^2+u^2}-u]/a$, $\lambda_v=[\sqrt{a^2+v^2}+v]/a$,
$u=t-\rho$, $v=t+\rho$, $\Xi = 1+\lambda_u\lambda_v +
2[(1-\alpha^{-2})\lambda_u\lambda_v]^{1/2}$ and a constant term in $m/f$
has been removed for simplicity as this cannot affect whether
(\ref{crit2}) is satisfied, though as is shown in section 3 this is
not an allowable transformation of an axisymmetric metric\footnote{
Equation (67)  of \ccite{CasMac90} contains a typographical error in
$\lambda_v$, but (68) is correct.}. At $\rho=0$, $u=v$, $\lambda_u\lambda_v=1$,
$\lambda_u+\lambda_v=2\sqrt{(a^2+t^2)}/a$, $\Xi$ is a constant which is
non-zero if $\alpha \neq 0$, and
$\lambda_u^2 + \lambda_v^2 = 2(a^2+2t^2)/a^2$, so the denominators of
$f$ and $m$ are
equal and non-zero, and the numerators are constant multiples of $a^2+2t^2$ and
$a^2+t^2$ respectively; thus the ratio $f/m$ at $\rho =0$ cannot be
constant as the criterion would demand (clearly in this case
$m \neq 0$). Hence $f$, $m$ and $\ell$ for
this metric cannot be linearly dependent. This is not surprising since
the metric is obtained by complex transformations from the Kerr
metric, for which $f$, $m$ and $\ell$ are linearly independent over
$\Real$ (e.g.\ take the formulae for them as functions of the
Boyer-Lindquist coordinates $r$ and $\theta$ as in equation (19.19) of
\ccite{KraSteMac80}) .

\section{Axisymmetry and global restrictions}

For a metric satisfying the conditions of the Proposition in the
previous section, the transformations used to bring it into its
canonical form work without difficulties when the $(\phi,\,X)$
coordinates can be considered to stretch to infinity in all
directions. However, as the notation itself suggests, one is often
interested in axisymmetric solutions in which the range of $\phi$ is
restricted to $[0,\,2\pi]$, and the $(\phi,\,X)$ surfaces thus have
cylindrical topology. I now consider the restrictions this imposes on
the permissible transformations.

First I consider the interpretation, in terms of the topological
identification, of the parameters of the transformations to the local
standard forms found in section 2. Let us consider transformations,
similar to equation (\ref{trans1}), of the form
\begin{equation}
\label{trans1a}
X = \hat{X} + \hat{\phi}/S, \qquad \phi^\prime = Q \hat{X} + \hat{\phi} .
\end{equation}
starting from coordinates $(\hat{\phi},\,\hat{X})$, for the moment
assumed to have infinite ranges, in which the metric takes the form
(\ref{diag}). (The following argument requires adaptation for the
other two cases but it is clear that the the main arguments about the
parameters needed to define a topological identification making the
$(\phi,\,X)$ plane into a cylinder hold in all cases. Note, however,
that if one starts from the form (\ref{nullKV}), exchange of
$\hat{\phi}$ and $\hat{X}$ may be required if the periodic coordinate
is null.)  Suppose that in these coordinates there is a topological
identification, making the $(\phi^\prime,\,X)$ plane into a cylinder,
in which $\phi^\prime$ is periodic with period $P^\prime$.

The identification of $(\phi^\prime,\,X)$ with
$(\phi^\prime+nP^\prime,\,X)$, for all integers $n$, identifies $(Q
\hat{X}+\hat{\phi},\, \hat{X}+\hat{\phi}/S)$ with $(Q
\hat{X}+\hat{\phi}+nP^\prime,\, \hat{X}+\hat{\phi}/S)$. Then the
origin is identified with $\hat{X}=-\hat{\phi}/S = nP^\prime/(S-Q)$
and the lines $Q \hat{X} + \hat{\phi}=0$ and $Q \hat{X} +
\hat{\phi}=-nP^\prime$ are identified. Thus two parameters ($S$ and $P
= SP^\prime/(S-Q)$ say) can be regarded as defining a point to be
identified with the origin, and the third ($Q$) as defining the lines
along which the identification is made.

We can in addition to (\ref{trans1a}) make a re-scaling of
$\phi^\prime$ to make its period $2\pi$, by
$\phi=2\pi\phi^\prime/P^\prime$, so the overall transformation is
\begin{equation}
\label{trans1b}
X = \hat{X} + \hat{\phi}/S, \qquad \phi = 2\pi(Q \hat{X} +
\hat{\phi})/P^\prime.
\end{equation}
We can relate this to the form (\ref{trans1}) by rescaling
$\hat{\phi}$ by $2\pi/P^\prime$ and setting $C=2\pi S/P^\prime$,
$B=2\pi Q/P^\prime$.

From this description we can see that the parameter $Q$ can be taken
to be zero, since a point $(\hat{\phi},\,\hat{X})$ is identified with
all points $(\hat{\phi}+kP,\,\hat{X}-kP/S)$ where $k$ is an integer,
and we can in particular describe this by using the lines $\hat{\phi}
= 0$ and $\hat{\phi} = kP$, and regarding $S$ as fixing the amount by which
points slip along those lines in the identification. The corresponding
coordinate transformation is
\begin{equation}
\label{trans2}
X = \hat{X} + \hat{\phi}/S, \qquad \phi = 2\pi\hat{\phi}/P.
\end{equation}

Thus if, as discussed above, the coordinates in the standard forms
have been scaled to remove as many inessential parameters as possible,
all distinct axisymmetric geometries locally equivalent to the
standard form can be obtained by applying a transformation
(\ref{trans2}) to the standard form (cf.\ \ccite{Sta82}) and
considering $\phi$ to have
period $2\pi$. The number of essential parameters in the axisymmetric
form is thus two more than the number in the corresponding locally
equivalent standard form. The same argument applies, {\em mutatis
mutandis}, if the standard form is itself taken to be axisymmetric:
the standard form will then contain an essential parameter, $\hat{P}$ say,
allowing the period of $\hat{\phi}$ to be $2\pi$, and distinct
stationary forms will be given by different values of $\hat{P}/P$.

Correspondingly, given an axisymmetric metric in our class,
we expect that transformations of the complementary form
\begin{equation}
\label{trans3}
\bar{X} = AX, \qquad \bar{\phi} = \phi + HX
\end{equation}
are allowed and are inessential in that they do not alter either the
local geometry or the topological identification. As this statement
disagrees with some earlier papers, the admissibility and properties
of a transformation (\ref{trans3}) will now be discussed in some
detail. I first note that it alters the surfaces on which $\phi$ is
constant, thus for example altering the $X$ axis, but it preserves the
surfaces on which $X$ is constant, including the $\phi$ axis. One way
to see directly that this transformation does not affect the geometry
is to note that it leaves the vector $\partial/\partial \phi$
invariant, i.e.\ preserves the uniquely-defined Killing  vector tangent to the
closed curves, which is not preserved, in general, by
(\ref{trans2}). To clarify the point further I consider allowable
coordinate changes directly.

Considering the angular coordinates $\phi$ and $\bar{\phi}$ to have a
range $2\pi$, the map (\ref{trans3}) from $\phi$ to $\bar{\phi}$ is
discontinuous at $\phi = 0$ (or $2\pi$) and similarly at $\bar{\phi} =
0$.  The transformation (\ref{trans3}) is therefore sometimes regarded
as inadmissible (see e.g.\ \ccite{SilHerPai95a}) or as destroying the
periodicity of $\phi$. However, periodic coordinates do not satisfy
the requirement that a coordinate chart in an $n$-dimensional
differential manifold should give a one-to-one map between an open set
of the manifold and an open set in $\Real^n$, and in my view the
discontinuities therefore merely reflect the fact that identification
of $0$ and $2\pi$ makes the coordinates improper at those points. To
construct a rigorous argument we should start by taking two or
more true coordinate patches; transformations will then be allowable if
after making them separately on each patch we can use the result to
construct a new (improper) system of coordinates $(\phi,\,X)$ with a periodic
$\phi$.

We note that for this to be possible the curves on which $\bar{X}$ is
constant must be the closed curves on which $X$ is constant, so,
ignoring changes of origin, we must take $\bar{X}$ proportional to
$X$, i.e.\ among homogeneous linear transformations in the
$(\phi,\,X)$ surfaces the first equation in (\ref{trans3}) is the most
general change in $X$ consistent with periodicity of $\phi$. Note that
although I agree with other authors (e.g.\ \ccite{Sta82,SilHerPai95a})
that a transformation $\bar{X} = AX+G\phi$ with $G \neq 0$ is
disallowed, I do so on the grounds of the global inadmissibility of
transforming $X$ in such a way as to alter the identification implied,
rather than the argument that the new $X$ would have a periodic
nature. This latter formulation does not take into account the fact
that the usual $\phi$, with 0 and $2\pi$ identified, is not strictly
an admissible coordinate, i.e.\ the periodic $X$ can only be derived
by using improper coordinates in the first place (a periodic time that
really was forced would of course be undesirable, except perhaps in
situations with closed timelike lines).

The transformation $\bar{\phi} = \phi + HX$ only redefines $\phi$
differently on each curve on which $X$ is constant, so one can see
intuitively why it is allowed by envisaging taking a stack of rings
and rotating each one by a different amount, which does not affect the
actual geometry at all.  An alternative way to describe the
transformation (\ref{trans3}) is that for both $\phi$ and $\bar{\phi}$
the range is infinite but $\phi$ and $\phi+2n\pi$ are identified for
any integer $n$ (and similarly for $\bar{\phi}$).

In fact, if two or more true coordinate patches had been used to cover the
manifold both before and after the transformation, each locally of the
canonical form, the transformations on the overlaps between new and
old coordinate regions would be smooth. For example, take patches
$U_1$ and $U_2$ defined respectively by $-3\pi/4 < \phi < 3\pi/4$ and
$\pi/4 < \phi < 7\pi/4$ and with $\phi$ coordinates denoted by
$\phi_1$ and $\phi_2$. The overlap $U_1 \cap U_2$ has two disjoint
parts, on one of which $\pi/4 < \phi_1 = \phi_2 < 3\pi/4$ and on the
other $5\pi/4 < \phi_2 = \phi_1+2\pi < 7\pi/4$. Now take a
transformation (\ref{trans3}) and use similar charts $V_1$ and $V_2$
defined by $\bar{\phi}$. The intersections, e.g.\ $V_1 \cap U_1$, each
consist of a countable infinity of mutually disjoint pieces which can
be labelled by the integer $n$ required so that the corresponding
transformation $\bar{\phi} = \phi + H X - 2\pi n$ gives values of
$\bar{\phi}$ in the appropriate range, e.g.\ $(-3\pi/4,\,3\pi/4)$  for $V_1$,
and this transformation is smooth on each piece. Thus this is an
acceptable coordinate change.

One can of course apply a general linear transformation with constant
coefficients to the ignorable coordinates in a true coordinate patch,
but this would not respect the importance of the global topology. For
example, the locally HSO KVs found in case (i) are not satisfactory as
globally HSO KVs unless $m=0$. If $m \neq 0$, the surfaces $\hat{\phi} = {\rm
constant}$ to which $\partial/\partial \hat{X}$ is orthogonal wind
round the axis in a helix, and the integral curves of the Killing
vector $\partial/\partial \hat{X}$ would meet such a surface
infinitely often. (Note that this last means that the surfaces
$\hat{X}=$ constant would not be achronal, despite being spacelike
everywhere.) The term `static' would normally be reserved for the case
where the surfaces pass through, and are orthogonal to, the axis and
meet each trajectory of $\partial/\partial X$ only once.

If (\ref{crit2}) is satisfied with real $B$ and $C$, then by a
transformation of the form (\ref{trans3}) with $H=-B$, we would have
$\bar{\ell} = \ell$ and $\bar{m}=m+B\ell$, and the linear dependence would
become $\bar{f} = (B-C)\bar{m}/\zeta$. This could then be related by a
transformation (\ref{trans2}) to a standard form (\ref{diag}). Similar
remarks apply to cases (ii) and (iii).

The parameters in (\ref{trans3}) can become essential if an additional
identification is made on the $X$ axis, i.e.\ in the toroidal case.
Such possibilities are increased if there is a third Killing vector
commuting with the first two, as in stationary cylindrically metrics
(where one can take $Y$ to be ignorable): then one can introduce
further complications by making identification(s) in one or more
variable(s) which identify the origin with a point in the
three-dimensional $(r,\,X,\,Y)$ space, and so on (cf.\
\ccite{Tod94}).

\section{Physical interpretation of the topological parameters}

The parameters $S$ and $P$ in the transformation (\ref{trans2}) do not
change the Riemannian curvature tensor and its covariant derivatives
at a given point, since locally they merely specify coordinate
transformations (though they will of course alter the coordinate
components of the tensors in the usual way). Hence they cannot affect
the values of the Cartan scalars obtained in the procedure for local
characterization of solutions of the Einstein equations
\cite{MacSke94,PaiRebMac93,SilHerPai95a}, though they will in general
alter the expressions for them in terms of the coordinates. However,
they are invariant topological characteristics of the axisymmetric
metric form, and it would be of interest to find a relation to
curvature. As Stachel noted (\citeyear{Sta82}), any such relation must
come from the global holonomy\footnote{Unless otherwise stated, I
consider only linear holonomy.} of the solution, by taking closed
curves around the axis, whose existence of course depends on the
identification and of which the simplest, and the only ones which are
trajectories of a Killing vector, are the curves on which $r$, $X$,
and $Y$ are constant and, using (\ref{trans2}), $\phi$ runs from 0 to
$2\pi$. In general the equations for parallel transport of a vector
$\bm{v}$ along these curves give a set of four coupled linear
homogeneous differential equations in the components of $\bm{v}$,
first-order with respect to $\phi$ and with coefficients independent
of $\phi$.

For definiteness, take the case $\zeta=-1$ in the Lewis form (with
$\mu=\nu=2V$). The parallel transport equations to
consider then take the form
\begin{equation}
\label{partrans}
 {{\d {\bf v} } \over {\d \phi}} = \bm{Av}
\end{equation}
where, if the components are given in terms of the coordinate numbering\\
$(x^1,\,x^2,\,x^3,\,x^4) = (r,\,z,\,\phi,\,t)$, the only non-zero
components of the matrix $\bm{A}$ (i.e.\ the relevant Christoffel
symbols $A^i{}_j = \{ {i \atop {j3}} \}$) are
\begin{equation}
\label{connection}
\begin{array}{ll}
A^1{}_3 =-\ell_{,r}/2\e^{2V}, &
A^1{}_4 = -m_{,r}/2\e^{2V}, \\
A^2{}_3 =-\ell_{,z}/2\e^{2V}, &
A^2{}_4 = -m_{,z}/2\e^{2V}, \\
A^3{}_1 = (f\ell_{,r}+mm_{,r})/2\rho ^2, &
A^3{}_2 = (f\ell_{,z}+mm_{,z})/2\rho ^2, \\
A^4{}_1 = (m\ell_{,r}-\ell m_{,r})/2\rho ^2, &
A^4{}_2 = (m\ell_{,z}-\ell m_{,z})/2\rho ^2.
\end{array}
\end{equation}

In general the matrix $\bm{A}$ will have four distinct eigenvectors
$\bm{v}_A$ with corresponding eigenvalues $\lambda_A$, and the general
solution of equation (\ref{partrans}) will be of the form \[
\bm{v}(\phi) = \sum_{A=1}^4 K_A \bm{v}_A \exp (\lambda_A \phi) \]
where the $K_A$ are arbitrary functions of $r$ and $z$: note that the
$\lambda_A$ and $\bm{v}_A$, although independent of $\phi$ (and $t$),
will also in general depend on $r$ and $z$.  The objective is to
examine how the parameters $S$ and $P$ affect the net change
$\bm{v}(2\pi) - \bm{v}(0)$ round a circle.

The eigenvalue equation for $\bm{A}$ takes the form
 \[ \lambda^4 + b_2 \lambda^2 + b_4 = 0 \]
where \[ b_4 = \det \bm{A} =
-(m_{,r}\ell_{,z}-m_{,z}\ell_{,r})^2/16\e^{4V}\rho ^2, \] \[ b_2 =
(g^{ab}\bm{w}^1_a \bm{w}^1_b +g^{ab}\bm{w}^2_a \bm{w}^2_b)/4\e^{2V}, \] and
the one-forms $\bm{w}^1$ and $\bm{w}^2$ lie in the $(\phi,\,t)$ plane
and have components $(\ell_{,r},\,m_{,r})$ and
$(\ell_{,z},\,m_{,z})$ respectively in that plane.
Thus the eigenvalues occur in pairs $\pm \sqrt{y}$ for $y$ satisfying $y^2+b_2
y + b_4=0$; non-trivial holonomy arises unless for each $\lambda_A$,
$\exp(2\lambda_A \pi) = 1$, and this would imply that each $\lambda_A =
in_A$ for some integer $n_A$, so in particular $b_4$ would be positive.

If the metric is in case (i) with $\zeta = -1$, the transformation
(\ref{trans2}) is made, with $P$ assumed to be positive, and $F$ and $L$ are
the $f$ and $\ell$ of the diagonal form, then $f = F$, $m = P F/2\pi
S$, and $\ell = P^2 (S^2 L-F)/4\pi^2 S^2$. Now the holonomy of the
transformed metric is given by
\[ b_2 = P^2[F(L_{,r}{}^2 + L_{,z}{}^2)-L(F_{,r}{}^2 +
F_{,z}{}^2)/S^2]/(16 \pi^2 \e^{2V} FL) \] and \[ b_4 =
-P^4(F_{,r}L_{,z}-F_{,z}L_{,r})^2/256 S^2 \pi^4 e^{4V} FL. \] It is
easily seen that for given $F$ and $L$, the values of $P$ and $S$ in
general affect the holonomy. All the eigenvalues scale with $P$, as
one would expect since this sets the scale of $\phi$ relative to
$\hat{\phi}$.  $S$ (or, more precisely, since a transformation
$\bar{X} = -X$ alters nothing essential in the holonomy, $|S|$), also
alters the set of eigenvalues: in particular two of the eigenvalues
are zero when $S=\infty$. When $S \neq \infty$, $b_4 < 0$ in general,
and in that case at least two of the eigenvalues are real and hence
cannot lead to $\exp (2\lambda_A \pi) = 1$. (The two additional
parameters of a general homogeneous linear transformation of
coordinates, which appear in (\ref{trans3}), cannot affect the holonomy
since they do not affect the invariant definition of the curves or of
parallel transport: however, as a check, an explicit computation was
performed to confirm this.)

There are of course exceptional cases where the parameters do not
affect the holonomy. If $F$ and $L$ are both constant, so that the
spacetime is 2$+$2 decomposable, all the eigenvalues are zero and the
holonomy is trivial. If the gradients of $F$ and $L$ are everywhere
parallel so that $F$ is a function of $L$ (e.g.\ if both depend on
only one of the coordinates $r$ and $z$), at least two of the
eigenvalues are zero. If in addition $F$ is constant, $S$ does not
affect the holonomy; this arises in the case of flat space with
identifications and in that case the linear holonomy depends only on
$P$ (the $\beta$ of \ccite{Tod94}), though $S$ (his $\alpha$ or $\gamma$) does
appear in the affine holonomy.

One can also see from the formulae above that, as one would expect, if
$\hat{\phi}$ is scaled up by a factor $K$, $P$ and $C^2$ scale up by the same
factor and $L$ by its inverse, while if $\hat{X}$ is scaled up by a
factor $A$, $C^2$ and $F$ scale by its inverse.

To illustrate this, consider the case (\ref{regionI}). There
are a pair of zero eigenvalues, with one eigenvector in the $z$
direction and another in a certain direction in the $(\phi,\,t)$
plane, and a pair of complex conjugate eigenvalues whose squares,
obtained with the help of REDUCE, are \[ -R \sinh (5\epsilon + \theta)
/(16 r \e^{2V} \cosh^4 \epsilon \sinh \epsilon).\] In these solutions
$\e^{2V} = (R/r)^{2a^2R^2} \exp(-a^2R^2)$. The values of $\epsilon$
and $R$ (or $a$ and $R$) are seen to affect the holonomy. One may
expect that van Stockum's third case leads to similar formulae with
trigonometric functions replacing the hyperbolic ones.

Although the additional parameters given by $A$ and $H$ in
(\ref{trans3}) are inessential, they may, as mentioned earlier, still
be required if the solution is matched to an interior using
Lichnerowicz's form of the matching conditions, see e.g.\
\ccite{BonMacSan97}. The interior of course need not obey
(\ref{crit2}) even when the exterior does. If on the other hand, a
metric covered by the earlier Proposition is continued to an axis, its
essential parameters may be interpreted as properties of line
masses. For example, \ccite{Lin85} considered static cylindrically
symmetric Einstein spaces and chose among them the ones interpretable
as cosmic strings: thus only a line mass on the axis appeared.

In \ccite{SilHerPai95a}, the parameters in the general vacuum solution
of case (i) above, of which the metric (\ref{regionI}) is a special
case, are considered. The metric was taken in the form
\begin{equation}
\label{weyl}
\begin{array}{lcl}
f &=& (a^2n^2-c^2r^{2n})/an^2r^{n-1} \\
m &=& -(a^2bn^2 + c(n-bc)r^{2n})/an^2r^{n-1} \\
\ell &=& (-a^2b^2n^2+(n-bc)^2r^{2n})/an^2r^{n-1},
\end{array}
\end{equation}
so $-f+((n-2bc)m+c\ell)/b(bc-n)=0$, and the roots of the quadratic
(\ref{quadr}) are $-1/b$ and $-c/(bc-n)$, which is consistent with
(3.7) and (3.8) of \ccite{SilHerPai95a}. The root $B = -c/(bc-n)=-H$
can be used in the transformation (\ref{trans3}), leading to a metric
of the same form with $\bar{a}=n^2a/(bc-n)^2$, $\bar{b} = b(bc-n)/n$,
$\bar{c}=0$ and $\bar{m}=\bar{b}\bar{f}$. The holonomy round the circles has
two zero eigenvalues, as noted above, and the non-zero eigenvalues are
the square roots of
\[ {{a^2 b^2 n^2 (n-1)^2 - (n+1)^2(bc-n)^2 r^{2n}} \over
{4an^2\e^V r^{n+1}}} \] where $\e^{4V} = r^{n^2+1}$. The value of
$\bar{f}$ gives as our comparison metric, without any rescaling of
$t$, the case with $F=n^2a/(bc-n)^2r^{n-1}$, $L=r^{n+1}$.
Then $P^2= 4\pi^2(bc-n)^2/n^2a$ and $C^2=a/b^2$. Scaling of
$\hat{X}$ to the perhaps more natural $F=1/r^{n-1}$ would scale $C^2$
to $(bc-n)^2/b^2n^2$.

By matching to a shell, \ccite{Sta84} identified (using my notation)
$P$ with an energy density, $n$ with stress-energy, and $C$ with the
rotation rate of the shell relative to a flat interior. Similarly,
\ccite{SilHerPai95a} interpreted $n$ as the Newtonian mass per unit
length (or the total mass of a fluid interior region). In the case
$c=0$ and $n=1$, they identified $a$ and $b$ as the energy density and
angular momentum of a string on the axis (cf.\ \ccite{JenKuc93}). By
matching to a fluid interior they interpreted $c$ as due to the
vorticity. However, this arises because in their treatment the
world-lines of the fluid flow in the interior invariantly define the
$t$ axis at the interface, and the use of admissible coordinates in
matching, in the Lichnerowicz sense, then leads to the appearance of
this quantity in form for the exterior space, although it is not an
invariant of the exterior space.

Although the above discussion is based only on the case (\ref{diag}),
similar results for holonomy will clearly hold in the other cases.
\ccite{SilHerPai95} correspondingly found similar results for
parameter identification by matching for the case (iii) metric
obtainable from the metric (\ref{weyl}) by taking $n$ to be pure
imaginary.

\section{Concluding remarks}

The proposition in section 2 shows that linear dependence over $\Real$
of the metric components in canonical coordinates for a metric with an
orthogonally-transitive commuting $G_2$ leads to (real or complex) HSO
KVs, or a null KV, and hence to local coordinate transformations to
one of the metric forms (\ref{diag}--\ref{windmill}). These
transformations contain in general four constant parameters (the
coefficients in a linear transformation of the canonical coordinates
with constant coefficients).  In axisymmetric metrics, two of these
parameters define the topological identification and can physically be
identified from holonomy round circles along which only $\phi$ varies,
and hence (assuming that $\rho$ is single-valued, so that the concept of the
interior is well-defined) with properties of the sources interior to a given
value of $\rho$, including possible line sources on an axis.

This leaves the difficulty that, as \ccite{MarSen95} have pointed out,
while one can develop a proper theory for regular axes (see
\ccite{WilCla96} for example), such a theory
for everywhere singular axes does not exist. In particular, one cannot
always attach a well-defined meaning to statements such as `the field
is axisymmetric about an infinite axis' \cite{KraSteMac80}.  Taking
$\ell >0$ at some $(r,\,Y)$ does not guarantee $\ell > 0$ for all
$(r,\,Y)$ but one usually wants to consider regions where $\ell >0$
and Mars and Senovilla argue that one should require $\ell>0$ in the
neighbourhood of an axis. One may note, for example, that an initially
diagonal form to which the transformation (\ref{trans2}) has been
applied may then have $\ell <0$ near the axis even if $L = 0$ there. In
practice fields which do not have regular axes are nevertheless
routinely described as cylindrically symmetric, and in particular even
those such solutions which if continued to an axis would not be
regular there may form part of a globally regular solution in which
they are exteriors for some regular interior with different
energy-momentum content: for example the solution (\ref{regionI}) is
an exterior for van Stockum's cylinders of rotating dust
\cite{Sto37,Bon80}.

This possibility means that rather than associating the parameters of
the topological identification with a line source on the axis, they
may be associated with the physical characteristics of the source
region to which the solution is matched, and if the matching is done
in the Lichnerowicz form, it may also fix some inessential parameters
of the exterior. It should be noted that the arguments of sections 3
and 4 can be extended, {\em mutatis mutandis}, to spacetimes not
obeying the Proposition in Section 2.

\section*{Acknowledgements}

The idea for the reformulation of the criterion of
\ccite{CasMac90} arose from considering the cylindrically symmetric
Einstein spaces in the form given by \ccite{San93}, with a view to
interpreting the parameters in those solutions: this work, which was
carried out during a visit to QMW by N.O. Santos supported by an
E.P.S.R.C. Visiting Fellowship award, will be reported on separately.
I am grateful to W.B. Bonnor and N.O. Santos for their stimulus and
for critical discussions which forced me to develop and clarify the
ideas presented here, and to E.P.S.R.C. for the grant. A second
stimulus was provided by a private communication from J.D. Barrow and
K.E. Kunze. J. Katz provided some useful details of the calculations
in \ccite{PirSafKat86}, M. Mars, J.M.M. Senovilla, J. Stachel and J.A. Vickers
made useful comments on a draft, B. Steadman checked and corrected some of the
calculations of holonomy and S.T.C. Siklos helped provide a reference.


\end{document}